\documentclass[12pt]{article}
\usepackage{graphicx}
\usepackage{graphics}
\title{Evaluation of some second moment and other integrals for the Riemann, Hurwitz, and Lerch zeta functions} 
\author{Mark W. Coffey\\
Department of Physics\\
Colorado School of Mines\\
Golden, CO  80401\\
(Received $\mbox{~~~~~~~~~~~~~~~~~~~~~~~~~~~~~~~2010}$)}
\date{January 15, 2011}
\begin{document}
\maketitle
\baselineskip=25 pt
\begin{abstract}

Several second moment and other integral evaluations for the Riemann zeta function 
$\zeta(s)$, Hurwitz zeta function $\zeta(s,a)$, and Lerch zeta function $\Phi(z,s,a)$ are presented.  Additional corollaries that are obtained include previously known special cases for the Riemann zeta function $\zeta(s)=\zeta(s,1)=\Phi(1,s,1)$.  An example special case is:
$$\int_R {{|\zeta(1/2+it)|^2} \over {t^2+1/4}}dt=2\pi[\ln(2\pi)-\gamma],$$  
with $\gamma$ the Euler constant.
The asymptotic forms of certain fractional part integrals, with and without logarithmic factors in the integrand, are presented.  Extensions and other approaches are mentioned.

\end{abstract}
 
\baselineskip=15pt
\centerline{\bf Key words and phrases}
\medskip 

\noindent

Hurwitz zeta function, Riemann zeta function, Dirichlet $L$ function, Lerch zeta function, second moment integral, fractional part integral, asymptotic form

\vfill
\centerline{\bf 2010 AMS codes} 
11M06, 11Y60, 11M35

\baselineskip=25pt
\pagebreak
\medskip
\centerline{\bf Introduction and statement of results}
\medskip

Let $\Phi(z,s,a)$ denote the Lerch zeta function \cite{sri}, $\zeta(s,a)=\Phi(1,s,a)$ the Hurwitz zeta function, and $\zeta(s)=\zeta(s,1)$ the Riemann zeta function 
\cite{edwards,ivic,riemann,titch}.  The function $\Phi$ may be analytically continued in
all three of its arguments, and satisfies the evident functional relation for integers $n$
$$\Phi(z,s,a)=z^n \Phi(z,s,n+a) +\sum_{k=0}^{n-1} {z^k \over {(k+a)^s}}.  \eqno(1.1)$$
We present several integral relations for a version of a second moment of $\zeta(s)$, $\zeta(s,a)$, and $\Phi(z,s,a)$.  Special cases of the latter results then give 
additional corollaries for the Riemann zeta function, and these include a result of Ivi\'{c} \cite{ivic} for the alternating form of the zeta function.  We illustrate that our results also extend to Dirichlet $L$ functions.
Throughout we write $\sigma=$ Re $s$ and $s=\sigma+it$. 

Moment integrals of the Riemann zeta function in particular have been of interest from
many points of view.  We recall that even the simple average of $|\zeta(s)|^2$ on vertical
lines in the right-half of the critical strip $0<\sigma<1$ has important application.
For Bohr and Landau based their theorem on the proportion of zeros within strips adjacent
to the critical line $\sigma=1/2$ by using the uniform boundedness of this average by $\zeta(2\sigma)$ for $\sigma>1/2$ (e.g., \cite{edwards} Ch. 9).

We recall the alternating Hurwitz zeta function, $\Phi(-1,s,a)$, valid for $\sigma>0$, when given by the series
$$\zeta_a(s,a)=\sum_{n=0}^\infty {{(-1)^n} \over {(n+a)^s}}
=2^{-s}\left[\zeta\left(s,{a \over 2}\right)-\zeta\left(s,{{a+1} \over 2}\right)\right].  \eqno(1.2)$$
Correspondingly, we have the special cases $\zeta(s,1/2)=(2^s-1)\zeta(s)$ and
$$\zeta_a(s,1)=\zeta_a(s)=(1-2^{1-s})\zeta(s).  \eqno(1.3)$$

In the following $(a)_j=\Gamma(a+j)/\Gamma(a)$ is the Pochhammer symbol.

We have
{\newline \bf Proposition 1}.  Let $\sigma>1$ and $a>0$.  Then we have (a)
$$\int_{-\infty}^\infty \left|{{\zeta(s,a)} \over s}\right|^2 dt={\pi \over \sigma}
[2\zeta(2\sigma-1,a)+(1-2a)\zeta(2\sigma,a)],  \eqno(1.4)$$
and (b)
$$\int_{-\infty}^\infty \left|{{\zeta'(s,a)} \over s}\right|^2 dt={\pi \over \sigma}
\left[\zeta''(2\sigma,a)+2\sum_{m=0}^\infty {{(\ln (a)_m)\ln(m+a)} \over {(m+a)^{2\sigma}}} \right].  $$
In particular, part (a) at $a=1/2$ yields
{\newline \bf Corollary 1}
$$\int_{-\infty}^\infty \left|{{(2^s-1)\zeta(s)} \over s}\right|^2 dt=2{\pi \over \sigma}
(2^{2\sigma-1}-1)\zeta(2\sigma-1),  \eqno(1.5)$$
and at $a=1$ gives
{\newline \bf Corollary 2}
$$\int_{-\infty}^\infty \left|{{\zeta(s)} \over s}\right|^2 dt={\pi \over \sigma}
[2\zeta(2\sigma-1)-\zeta(2\sigma)].  \eqno(1.6)$$
This latter relation is proved directly by two different methods in the Appendix.  

{\bf Proposition 2}.  Let $\sigma>0$ and $a>0$.  Then we have
$$\int_{-\infty}^\infty \left|{{\zeta_a(s,a)} \over s}\right|^2 dt={\pi \over \sigma}
\zeta_a(2\sigma,a).  \eqno(1.7)$$
Therefore, we have
{\newline \bf Corollary 3}.  We have for $\sigma>0$,
$$\int_{-\infty}^\infty \left|{{\zeta_a(s)} \over s}\right|^2 dt={\pi \over \sigma}
\zeta_a(2\sigma),  \eqno(1.8)$$
being Theorem 1 of \cite{ivic}.
We recall that
$$\zeta_a(1,a)={1 \over 2}\left[\psi\left({{1+a} \over 2}\right)-\psi\left({a \over 2}\right)\right], \eqno(1.9)$$
where $\psi=\Gamma'/\Gamma$ is the digamma function \cite{nbs,andrews,grad}.  We then obtain
{\newline \bf Corollary 4}.  For Re $a>0$ we have
$$\int_{-\infty}^\infty \left|{{\zeta_a(1/2+it,a)} \over {1/2+it}}\right|^2 dt
=\pi \left[\psi\left({{1+a} \over 2}\right)-\psi\left({a \over 2}\right)\right]. \eqno(1.10)$$
In turn, $\zeta_a(1,1)=\ln 2$, and we recover the case of Corollary 1 of \cite{ivic}.
Since we have the reflection formula $\psi(1-z)-\psi(z)=\pi \cot \pi z$, we obtain
with $a=1/2$ in (1.10) 
{\newline \bf Corollary 5}.  We have
$$\int_{-\infty}^\infty \left|{{\zeta_a(1/2+it,1/2)} \over {1/2+it}}\right|^2 dt
=\pi^2. \eqno(1.11)$$
{\newline \bf Corollary 6}.  We have
$$\int_{-\infty}^\infty \left|{{\zeta_a(1/2+it,2)} \over {1/2+it}}\right|^2 dt
=2\pi(1-\ln 2), \eqno(1.12)$$
where $\zeta_a(s,2)=1-\zeta(s)+2^{1-s}\zeta(s)$.

Other identities follow from specializations in \cite{ivic}.  Let the function $\varphi(x)$
be defined as in (3) of \cite{ivic}:  with $\chi_A(x)$ the characteristic function of the set
$A$,
$$\varphi(x)=\sum_{m=1}^\infty \sum_{n=1}^\infty \int_1^x \chi_{[2m-1,2m)}\left({x \over u}
\right)\chi_{[2n-1,2n)}(u){{du} \over u}, ~~~~~~~x \geq 1.  \eqno(1.13)$$  
\newline{\bf Corollary 7}.  We have (a)
$$\int_1^\infty {{\varphi(x)} \over x^2}dx=\ln^2 2,  \eqno(1.14)$$
(b) 
$$\int_0^\infty [1-\cos(t \ln 2)] |\zeta(1+it)|^2 {{dt} \over {1+t^2}}={\pi^3 \over {48}},
\eqno(1.15)$$
and
(c)
$$\int_0^\infty [1-\cos(t \ln 2)]^2 |\zeta(1+it)|^4 {{dt} \over {(1+t^2)^2}}
=\int_1^\infty {{\varphi^2(x)} \over x^3}dx.  \eqno(1.16)$$
\newline{\bf Conjecture 1}.  
$$\lim_{s \to 0} s^2 \int_1^\infty {{\varphi(x)} \over x^{s+1}}dx ={1 \over 4}.  \eqno(1.17)$$

{\bf Proposition 3}.  Let $|z| \leq 1$, $a>0$, and $\sigma>0$, unless $z=1$, in 
which case $\sigma>1$.  Then
$$\int_{-\infty}^\infty \left|{{\Phi(z,s,a)} \over s}\right|^2 dt={\pi \over \sigma}
{1 \over {(z-1)}}[(z+1)\Phi(|z|^2,2\sigma,a)-2\Phi(z^*,2\sigma,a)].  \eqno(1.18)$$
We note that this Proposition recovers Proposition 2 when $z=-1$.    

Let $\mu(n)$ be the M\"{o}bius function and $d(n)$ the number of divisors of $n$.
\newline{\bf Proposition 4}.  We have for $\sigma>1$ (a)
$$\int_{-\infty}^\infty {1 \over {|\zeta(s)|^2}}{{dt} \over {|s|^2}}={\pi \over \sigma}
\left[{{\zeta(2\sigma)} \over {\zeta(4\sigma)}}+2\sum_{m=1}^\infty {{\mu(m)} \over m^{2
\sigma}} \sum_{n=1}^{m-1} \mu(n) \right],  \eqno(1.19)$$    
and (b)
$$\int_{-\infty}^\infty {1 \over {|\zeta(s)|^2}}{{dt} \over {|s|^2}}
\leq \int_{-\infty}^\infty \left|{{\zeta(s)} \over s}\right|^2 dt.  \eqno(1.20)$$
(c) For $\sigma>1$ we have
$$\int_{-\infty}^\infty {{|\zeta^2(s)|^2} \over {|s|^2}}dt={\pi \over \sigma}
\left[{{\zeta^4(2\sigma)} \over {\zeta(4\sigma)}}+2\sum_{m=1}^\infty {{d(m)} \over m^{2
\sigma}} \sum_{n=1}^{m-1} d(n) \right].  \eqno(1.21)$$

Let $L(s,\chi)$ be the Dirichlet $L$ function corresponding to Dirichlet character $\chi$
modulo $k$ and $\chi_0$ the principal character.
{\newline \bf Proposition 5}.  For $\chi$ a nonprincipal character let $\sigma>0$, and
otherwise take $\sigma>1$.  Then we have
$$\int_{-\infty}^\infty \left|{{L(s,\chi)} \over s}\right|^2 dt={\pi \over \sigma}
\left[L(2\sigma,\chi_0)+ 2\sum_{m=1}^\infty \sum_{n=1}^m {{\chi(n)\chi^*(m)} \over m^{2\sigma}}\right].  \eqno(1.22)$$

{\bf Proposition 6}.  Let $\sigma>1$ and $a>0$.  Then we have (a)
$$\int_{-\infty}^\infty {{|\zeta(s,a)|^2} \over {|s|^4}}dt={\pi \over \sigma^2}\left[
{1 \over \sigma}\zeta(2\sigma-1,a)+{1 \over \sigma}\left({1 \over 2}-a\right)\zeta(2\sigma,a)
-\zeta'(2\sigma-1,a)+a\zeta'(2\sigma,a)\right.$$
$$\left. -\sum_{m=0}^\infty {{\ln (a)_m} \over {(m+a)^{2\sigma}}}\right], \eqno(1.23)$$
and (b)
$$\int_{-\infty}^\infty {{|\zeta(s)|^2} \over {|s|^4}}dt={\pi \over \sigma^2}\left[
{1 \over \sigma}\zeta(2\sigma-1)-{1 \over {2\sigma}}\zeta(2\sigma)
-\zeta'(2\sigma-1)+\zeta'(2\sigma)-\sum_{m=1}^\infty {{\ln (m-1)!} \over m^{2\sigma}}\right]. \eqno(1.24)$$
Let $\sigma>0$ and $a>0$.  Then we have (c)
$$\int_{-\infty}^\infty {{|\zeta_a(s,a)|^2} \over {|s|^4}}dt={\pi \over {2\sigma^2}}\left[
{1 \over \sigma}\zeta_a(2\sigma,a)-\zeta_a'(2\sigma,a)+\zeta'(2\sigma,a)-2\sum_{m=0}^\infty
\sum_{n=0}^{m-1} {{(-1)^{m+n}} \over {(m+a)^{2\sigma}}}\ln(n+a)\right].  \eqno(1.25)$$

In \cite{ivic2} (Corollary 1), Ivi\'c has given the special case relation for $0<\sigma<1$
$${1 \over {2\pi}}\int_{-\infty}^\infty {{|\zeta(\sigma+it)|^2} \over {\sigma^2+t^2}}dt
=\int_0^\infty {{\{x\}^2} \over x^{2\sigma+1}}dx, \eqno(1.26)$$
where $\{x\}=x-[x]$ denotes the fractional part of $x$.  Based upon (1.26) we provide the following explicit evaluation.
{\newline \bf Proposition 7}.  For $0<\sigma<1$ we have
$${1 \over {2\pi}}\int_{-\infty}^\infty {{|\zeta(\sigma+it)|^2} \over {\sigma^2+t^2}}dt
=-{1 \over \sigma}\left[{{\zeta(2\sigma)} \over 2}+{{\zeta(2\sigma-1)} \over {2\sigma-1}}
\right].  \eqno(1.27)$$
From this result, we may draw other conclusions pertinent to the critical line.  For instance,
with $\gamma=-\psi(1)$ the Euler constant, we have
\newline{\bf Corollary 8}.  We have (a)
$$\int_R {{|\zeta(1/2+it)|^2} \over {t^2+1/4}}dt=2\pi[\ln(2\pi)-\gamma], \eqno(1.28a)$$
and (b)
$$\int_R \cos(t\ln 2){{|\zeta(1/2+it)|^2} \over {t^2+1/4}}dt={\pi\over \sqrt{2}} [2\ln 2 +3\ln \pi-3\gamma]. \eqno(1.28b)$$
We may remark that the right side of (1.27) confirms that the quantity in brackets is negative, and in fact, since $\zeta(0)=B_1=-1/2$ and $\zeta(-1)=-B_2/2=-1/12$, where $B_j$ is the $j$th Bernoulli number, this sum ranges from $-1/6$ at $\sigma=0$ to $-\infty$ at $\sigma=1$.

Also given in \cite{ivic2} (Theorem 3) is the following asymptotic formula.  Let $M \geq 1$
be a fixed integer.  Then as $T \to \infty$
$$\int_0^\infty {{\{x\}} \over x}e^{-x/T}dx={1 \over 2}\ln T-{\gamma \over 2}+{1 \over 2}
\ln(2\pi)$$
$$+\sum_{m=1}^M {{\zeta(1-2m)} \over {(2m-1)!(1-2m)}}T^{1-2m}+O_M\left(T^{-2M-1}\right).  
\eqno(1.29)$$
Ivi\'{c} used contour integration to obtain this formula.  We show an alternative route,
using real variable asymptotics.  While successive higher order terms may be systematically
found, we are content to state the following.
{\newline \bf Proposition 8}.  (a) For $T \to \infty$ 
$$\int_0^\infty {{\{x\}} \over x}e^{-x/T}dx={1 \over 2}\left[\ln T-\gamma+
\ln(2\pi)\right]+{1 \over {12T}} +O\left({1 \over T^3}\right).  \eqno(1.30)$$
(b) For $T \to \infty$
$$\int_0^\infty {{\{x\}^2} \over x^2}e^{-x/T}dx=\ln(2\pi)-\gamma+[-2+2\gamma+12\ln A
-\ln (2\pi)^3-2\ln T]{1 \over {6T}}$$
$$-{1 \over {24T^2}}+{1 \over {2160T^3}}+O\left({1 \over T^4}\right),  \eqno(1.31)$$
where the Glaisher constant $A$ is such that $\ln A=1/12-\zeta'(-1)$.
While our approach may not as easily obtain the higher order terms in case (a), it
applies to a wider class of integrands as illustrated by part (b).

Let $\varphi_1(x)=\{x\}$, and recursively define functions
$$\varphi_n(x)=\int_0^\infty \{u\}\varphi_{n-1}\left({x \over u}\right){{du} \over u}, 
~~n \geq 2.  \eqno(1.32)$$
Then we have the following sharpening of (8) of Theorem 1 of \cite{ivic2}.
\newline{\bf Proposition 9}.  For $x>1$ we have
$$\varphi_2(x)={1 \over 4}\ln x + {1 \over 2}\ln (2\pi)+o(1), \eqno(1.33)$$
and
$$\varphi_n(x)={1 \over {2^n (n-1)!}}\ln^{n-1} x + {n \over{2^n(n-2)!}}\ln (2\pi)\ln^{n-2}x
+ O(\ln^{n-3}x).  \eqno(1.34)$$ 

Special values of the function $\varphi_2$ may be found exactly at rational and integer
argument.  As a sample of this, we have
{\newline \bf Proposition 10}.  We have $\varphi_2(1)=2(1-\gamma)$, 
$\varphi_2(2)=4(1-\gamma)-\ln 2$, $\varphi_2(3)=6(1-\gamma)-2\ln 2$, 
$\varphi_2(4)=8(1-\gamma)-2\ln 3$, $\varphi_2(5)=10(1-\gamma)-\ln (2304/125)$, 
$\varphi_2(6)=12(1-\gamma)-2\ln (20/3)$, $\varphi_2(7)=14(1-\gamma)-2\ln (8640)+7\ln 7$,
$\varphi_2(8)=16(1-\gamma)-2\ln (945/64)$, $\varphi_2(9)=18(1-\gamma)-2\ln (143360/6561)$,
and $\varphi_2(10)=20(1-\gamma)+11\ln(5/2)-2\ln(5103)$.

We may remark that $\varphi_2$ has the following summatory representation:
$$\varphi_2(x)=\sum_{j=0}^\infty \int_j^{j+1} (u-j)\left\{{x \over u}\right\} {{du} \over u}
=\sum_{j=0}^\infty \int_{1/(j+1)}^{1/j} \left({1 \over y^2}-{j \over y}\right)\{ xy \} dy.
\eqno(1.35)$$

More general moment integrals of the form 
$$\int_{-\infty}^\infty {{|\zeta(s,a)|^{2m}} \over {|s|^2}} dt \eqno(1.36)$$
have been investigated.  Bernard has written a compact form of these integrals, and
made a detailed study of the $m=2$ case \cite{bunpub}.  Although the $m=2$ evaluation, as
expected, contains terms such as $\zeta(4\sigma,a)$, $\zeta^2(2\sigma,a)$, $\zeta(4\sigma-1,a
)$, and further products $\zeta(2\sigma+b,a)\zeta(2\sigma+c,a+3)$ with $b$, $c=0$, $1$, 
the complete lengthly expression contains, for instance, complicated triple summations.

\medskip
\centerline{\bf Proof of Propositions}

We shall have recourse to the integral for $\sigma>0$ and real values of $\alpha$
$$\int_{-\infty}^\infty {{\cos \alpha x} \over {\sigma^2+x^2}}dx={\pi \over \sigma}
e^{-|\alpha| \sigma}.  \eqno(2.1)$$

{\it \bf Proposition 1}.  (a) For $\sigma>1$ we have
$$\int_{-\infty}^\infty \left|{{\zeta(s,a)} \over s}\right|^2 dt=\int_{-\infty}^\infty
\left|\sum_{n=0}^\infty {1 \over {(n+a)^s}}\right|^2 {{dt} \over {\sigma^2+t^2}}$$
$$=\sum_{n,m \geq 0}{1 \over {[(n+a)(m+a)]^\sigma}}\int_{-\infty}^\infty \left({{m+a}
\over {n+a}}\right)^{it} {{dt} \over {\sigma^2+t^2}}$$
$$=\sum_{n,m \geq 0}{1 \over {[(n+a)(m+a)]^\sigma}}\int_{-\infty}^\infty \cos\left[t\ \ln \left({{m+a} \over {n+a}}\right)\right] {{dt} \over {\sigma^2+t^2}}$$
$$={\pi \over \sigma}\zeta(2\sigma,a)+2 \sum_{m=0}^\infty \sum_{n<m} {1 \over {[(n+a)(m+a)]^\sigma}}\int_{-\infty}^\infty \cos\left[t\ \ln \left({{m+a} \over {n+a}}\right)\right] {{dt} \over {\sigma^2+t^2}}$$
$$={\pi \over \sigma}\left[\zeta(2\sigma,a)+2 \sum_{m=0}^\infty \sum_{n=0}^{m-1}
{1 \over {(m+a)^{2\sigma}}}\right]$$
$$={\pi \over \sigma}\left[\zeta(2\sigma,a)+2 \sum_{m=0}^\infty {m \over {(m+a)^{2\sigma}}}\right]$$
$$={\pi \over \sigma}\left\{\zeta(2\sigma,a)+2[\zeta(2\sigma-1,a)-a\zeta(2\sigma,a)]
\right\}$$
$$={\pi \over \sigma}[2\zeta(2\sigma-1,a)+(1-2a)\zeta(2\sigma,a)].  \eqno(2.2)$$
The series used for $\zeta(s,a)$ being absolutely convergent, the initial interchange
of summation and integration is justified.  Part (b) is proved similarly.

{\it \bf Proposition 2}.  For $\sigma>0$ we have
$$\int_{-\infty}^\infty \left|{{\zeta_a(s,a)} \over s}\right|^2 dt=\int_{-\infty}^\infty
\left|\sum_{n=0}^\infty {{(-1)^n} \over {(n+a)^s}}\right|^2 {{dt} \over {\sigma^2+t^2}}$$
$$=\sum_{n,m \geq 0}{{(-1)^{m+n}} \over {[(n+a)(m+a)]^\sigma}}\int_{-\infty}^\infty \left({{m+a} \over {n+a}}\right)^{it} {{dt} \over {\sigma^2+t^2}}$$
$$=\sum_{n,m \geq 0}{{(-1)^{m+n}} \over {[(n+a)(m+a)]^\sigma}}\int_{-\infty}^\infty \cos\left[t\ \ln \left({{m+a} \over {n+a}}\right)\right] {{dt} \over {\sigma^2+t^2}}$$
$$={\pi \over \sigma}\zeta(2\sigma,a)+2 \sum_{m=0}^\infty \sum_{n<m} {{(-1)^{m+n}} \over {[(n+a)(m+a)]^\sigma}}\int_{-\infty}^\infty \cos\left[t\ \ln \left({{m+a} \over {n+a}}\right)\right] {{dt} \over {\sigma^2+t^2}}$$
$$={\pi \over \sigma}\left[\zeta(2\sigma,a)+2 \sum_{m=0}^\infty \sum_{n=0}^{m-1}
{{(-1)^{m+n}} \over {(m+a)^{2\sigma}}}\right]$$
$$={\pi \over \sigma}\left[\zeta(2\sigma,a) + \sum_{m=0}^\infty {{(-1)^m} \over {(m+a)^{2\sigma}}}[1-(-1)^m]\right]$$
$$={\pi \over \sigma}[\zeta(2\sigma,a)+\zeta_a(2\sigma,a)-\zeta(2\sigma,a)]={\pi \over \sigma}\zeta_a(2\sigma,a).  \eqno(2.3)$$


{\it Corollary 7}.  (a) follows from the limit
$$\lim_{s\to 1} (1-2^{1-s})^2 \zeta^2(s) = \ln^2 2,$$
together with relation (4) in \cite{ivic}.  (b) follows by taking $\sigma=1$ in (1) in
\cite{ivic}.  (c) follows by Parseval's theorem for Mellin transforms and taking $\sigma=1$
in (4) in \cite{ivic}.  

Conjecture 1 is based upon relation (4) in \cite{ivic} and the fact that $\zeta^2(0)=1/4$.
The integral itself of the left side of (1.17) is slowly diverging, but we conjecture 
that the presented limit exists.

{\it \bf Proposition 3}.  We proceed as in the proof of Proposition 2, obtaining
$$\int_{-\infty}^\infty \left|{{\Phi(z,s,a)} \over s}\right|^2 dt
=\sum_{n,m \geq 0}{{z^n (z^*)^m} \over {[(n+a)(m+a)]^\sigma}}\int_{-\infty}^\infty \cos\left[t\ \ln \left({{m+a} \over {n+a}}\right)\right] {{dt} \over {\sigma^2+t^2}}$$
$$={\pi \over \sigma}\left[\sum_{n=0}^\infty {{|z|^2} \over {(n+a)^{2\sigma}}}
+2\sum_{m=0}^\infty \sum_{n=0}^{m-1} {{z^n (z^*)^m} \over {(m+a)^{2\sigma}}}\right]$$
$$={\pi \over \sigma}\left[\Phi(|z|^2,2\sigma,a)+2\sum_{m=0}^\infty {{(z^*)^m} \over
{(m+a)^{2\sigma}}} {{(1-z^m)} \over {(1-z)}}\right]$$
$$={\pi \over \sigma}\left[\Phi(|z|^2,2\sigma,a)+{2 \over {1-z}}[\Phi(z^*,2\sigma,a)
-\Phi(|z|^2,2\sigma,a)]\right].  \eqno(2.4)$$

{\it Remark}.  We have recently derived an integral representation for $\Phi$ that
generalizes that of (A.2) for $\zeta(s,a)$ \cite{coffey10} (Proposition 1).  With $P_1(x)=\{x\}-1/2$, $\{x\}$ the fractional part of $x$, for $a \in C/\{-1,-2,\ldots\}$, 
$s \in C$ when $|z|<1$, and $\sigma>1$ when $|z|=1$, we have
$$\Phi(z,s,a)={1 \over a^s}+{z \over {2(a+1)^s}}+\int_1^\infty {z^x \over {(x+a)^s}
}dx+\int_1^\infty \left[{{z^x \ln z} \over {(x+a)^s}}-{{sz^x} \over {(x+a)^{s+1}}}
\right]P_1(x)dx.  \eqno(2.5)$$
By following the second method of integral evaluation shown in the Appendix, this
$P_1$-based integral representation can also be used to prove Proposition 3.

{\it \bf Proposition 4}.  We use the Dirichlet series valid for $\sigma>1$
$$\sum_{n=1}^\infty {{\mu(n)} \over n^s}={1 \over {\zeta(s)}}, ~~~~~~
\sum_{n=1}^\infty {{\mu^2(n)} \over n^s}=\sum_{n=1}^\infty {{|\mu(n)|} \over n^s}={{\zeta(s)} \over {\zeta(2s)}}, \eqno(2.6)$$
with the latter corresponding to that for square-free numbers \cite{ivic} (pp. 32-33). Then
for part (a) we have
$$\int_{-\infty}^\infty {1 \over {|\zeta(s)|^2}}{{dt} \over {|s|^2}}=\int_{-\infty}^\infty
\left|\sum_{n=1}^\infty {{\mu(n)} \over n^{\sigma+it}}\right|^2 {{dt} \over {\sigma^2+t^2}}$$
$$=\sum_{n,m \geq 1}{{\mu(n)\mu(m)} \over {(nm)^\sigma}}\int_{-\infty}^\infty \left({m \over n} \right)^{it} {{dt} \over {\sigma^2+t^2}}$$
$$=\sum_{n,m \geq 1}{{\mu(n)\mu(m)} \over {(nm)^\sigma}}\int_{-\infty}^\infty \cos\left[t\ \ln \left({m \over n}\right)\right] {{dt} \over {\sigma^2+t^2}}$$
$$={\pi \over \sigma}\sum_{n=1}^\infty {{\mu^2(n)} \over n^{2\sigma}}+\sum_{n \geq 1} \sum_{\stackrel{m \geq 1}{m \neq n}}{{\mu(n)\mu(m)} \over {(nm)^\sigma}}\int_{-\infty}^\infty \cos\left[t\ \ln \left({m \over n} \right)\right] {{dt} \over {\sigma^2+t^2}}$$
$$={\pi \over \sigma}\left[{{\zeta(2\sigma)} \over {\zeta(4\sigma)}}+2 \sum_{m=1}^\infty \sum_{n=1}^{m-1} \mu(n) {{\mu(m)} \over m^{2\sigma}}\right].  \eqno(2.7)$$

For part (b), one way to proceed is to apply the triangle inequality to the right side of
(1.14), giving
$$\int_{-\infty}^\infty {1 \over {|\zeta(s)|^2}}{{dt} \over {|s|^2}} \leq 
{\pi \over \sigma}\left[{{\zeta(2\sigma)} \over {\zeta(4\sigma)}}+2[\zeta(2\sigma-1) -\zeta(2\sigma)]\right]$$
$$\leq {\pi \over \sigma}[\zeta(2\sigma)+2[\zeta(2\sigma-1) -\zeta(2\sigma)]]
={\pi \over \sigma}[2\zeta(2\sigma-1) -\zeta(2\sigma)].  \eqno(2.8)$$
The right member here being the integral of Corollary 2, part (b) is completed.

For part (c) we proceed as in (a), using the Dirichlet series for $\sigma>1$ (e.g., 
\cite{titch}, pp. 4-5)
$$\sum_{n=1}^\infty {{d(n)} \over n^s}=\zeta^2(s), ~~~~~~~~~~
\sum_{n=1}^\infty {{d^2(n)} \over n^s}={{\zeta^4(s)} \over {\zeta(2s)}}. \eqno(2.9)$$

{\it \bf Proposition 5}.  For an arbitrary convergent Dirichlet series 
$f(s)=\sum_{n=1}^\infty a_n n^{-s}$ we find from (2.1) that
$$\int_{-\infty}^\infty \left|{{f(s)} \over s}\right|^2 dt={\pi \over \sigma}\left[
\sum_{n=1}^\infty {{|a_n|^2} \over n^{2\sigma}} +2\sum_{m=1}^\infty \sum_{n=1}^{m-1}
{{a_na_m^*} \over m^{2\sigma}} \right].  \eqno(2.10)$$
When $a_n=\chi(n)$, $|\chi|^2$ is simply $0$ or $1$ according to $|\chi|^2=\chi_0$, and
the result follows.

{\it Remark}.  
Alternatively, the $L$ function may be written as a linear combination of Hurwitz zeta functions.  For $\chi$ a principal (nonprincipal) character and $\sigma > 1$ ($\sigma > 0$) 
we have
$$L(s,\chi) = \sum_{n=1}^\infty {{\chi(n)} \over n^s} ={1 \over k^s}\sum_{m=1}^k 
\chi(m) \zeta\left(s,{m \over k}\right).  \eqno(2.11)$$

{\it \bf Proposition 6}.  By acting with $-\partial/\partial \sigma$ on (2.1), 
for $\sigma>0$ and real values of $\alpha$ we have
$$\int_{-\infty}^\infty {{\cos \alpha x} \over {(\sigma^2+x^2)^2}}dx={\pi \over {2\sigma^2}}
\left(|\alpha|+{1 \over \sigma}\right)e^{-|\alpha| \sigma}.  \eqno(2.12)$$
(a) We proceed similarly to Proposition 1, but using (2.9) and 
$$\zeta'(s,a)=-\sum_{n=0}^\infty {{\ln(n+a)} \over {(n+a)^s}}.  \eqno(2.13)$$
(b) is the $a=1$ reduction of part (a), with $(1)_m=m!$.  For (c) we proceed similarly to
Proposition 2 and use
$$\zeta_a'(s,a)=-\sum_{n=0}^\infty (-1)^n{{\ln(n+a)} \over {(n+a)^s}}.  \eqno(2.14)$$

{\it Remark}.  By using integral representations for $\ln \Gamma$, one may write integral
representations for the last terms on the right sides of (1.23) and (1.24).  Stirling's
formula may be used to study the asymptotic behaviour of the summands there.  In (1.25) one
may write the inner sum $\sum_{n=0}^{m-1} (-1)^n \ln (n+a)=\sum_{\ell=0}^{[m/2]} \ln \left(
{{2\ell+a} \over {2\ell+a+1}}\right)$.

{\it \bf Proposition 7}.  We first have
{\newline \bf Lemma 1}.  Let $f$ be an integrable function, and $\psi^{(j)}$ the 
polygamma function (e.g., \cite{nbs,grad}).  Then (a)
$$\int_1^\infty f(\{x\}){{dx} \over x^\lambda}=\int_0^1 f(y)\zeta(\lambda,y+1)dy,
\eqno(2.15)$$
and (b) for integers $k\geq 0$
$$\int_1^\infty f(\{x\}){{dx} \over x^k}={{(-1)^k} \over {(k-1)!}}\int_0^1 f(y)
\psi^{(k-1)}(y+1)dy.  \eqno(2.16)$$
The relations (2.15) and (2.16) hold when the corresponding integrals converge.

{\it Proof}.  We have 
$$\int_1^\infty f(\{x\}){{dx} \over x^\lambda}=\sum_{\ell=1}^\infty \int_\ell^{\ell+1}
f(\{x\}){{dx} \over x^\lambda}$$
$$=\sum_{\ell=1}^\infty \int_0^1 {{f(y)} \over {(y+\ell)^\lambda}}dy
=\int_0^1 f(y)\zeta(\lambda,y+1)dy$$
$$=\int_0^1 f(y)\left[\zeta(\lambda,y)-{1 \over y^\lambda}\right]dy.  \eqno(2.17)$$
For part (b) we use the relation
$$\zeta(n+1,x)={{(-1)^{n+1}} \over {n!}} \psi^{(n)}(x).  \eqno(2.18)$$
Part (b) also takes the form
$$\int_1^\infty f(\{x\}){{dx} \over x^k}={{(-1)^k} \over {(k-1)!}}\int_0^1 f(y)
\left[\psi^{(k-1)}(y)-{{(-1)^{k-1}(k-1)!} \over y^k}\right]dy,  \eqno(2.19)$$
due to the functional equation of the polygamma function.

We now apply the Lemma with $f(u)=u^2$ to (1.21) and integrate by parts twice:
$${1 \over {2\pi}}\int_{-\infty}^\infty {{|\zeta(\sigma+it)|^2} \over {\sigma^2+t^2}}dt
=\int_0^\infty {{\{x\}^2} \over x^{2\sigma+1}}dx$$
$$=\int_0^1 y^2 \zeta(2\sigma+1,y)dy$$
$$={1 \over \sigma}\int_0^1 y \zeta(2\sigma,y)dy-{1 \over {2\sigma}}\zeta(2\sigma)$$
$$=-{1 \over \sigma}\left[{{\zeta(2\sigma)} \over 2}+{{\zeta(2\sigma-1)} \over {2\sigma-1}}
\right].  \eqno(2.20)$$
Herein, we have used the properties
$$\zeta(s+1,y)=-{1 \over s} {\partial \over {\partial y}}\zeta(s,y), ~~~~~~
\int_0^1 \zeta(s,y)dy=0 ~~\mbox{for} ~~\sigma<1.  \eqno(2.21)$$

For Corollary 8(a), we take the limit of (1.27) as $\sigma \to 1/2$ by using the Laurent
expansions of $\zeta(2\sigma)$ and $\zeta(2\sigma-1)$ about $\sigma=1/2$,
$$\zeta(2\sigma)={1 \over {2\sigma-1}}+\gamma +O\left(s-{1 \over 2}\right), \eqno(2.22a)$$
and
$$2{{\zeta(2\sigma-1)} \over {2\sigma-1}}=-{1 \over {2\sigma-1}}-\ln(2\pi) +O\left(s-{1 \over 2}\right). \eqno(2.22b)$$
These expansions are easily developed, keeping in mind the value $2\zeta'(0)=-\ln(2\pi)$
for (2.22b).  For (2.22a), the standard expansion in terms of Stieltjes constants 
$\gamma_k$, with $\gamma_0=\gamma$, may be employed (e.g., \cite{coffeyjmaa}).  
For part (b) we subtract a combination of the integral of Corollary 1 of \cite{ivic} and
that of part (a).

{\it Remarks}.  Relation (1.26) is the $n=1$ case of \cite{ivic2} (Theorem 2)
$${{\zeta^n(s)} \over {(-s)^n}}=\int_0^\infty \varphi_n(x)x^{-s-1}dx, ~~~~0<\sigma<1, \eqno(2.23)$$
with $\varphi_n(x)$ given in (1.32).
The moment integrals 
$${1 \over {2\pi}}\int_R \left|{{\zeta(s)} \over s}\right|^{2n}dt=\int_0^\infty \varphi_n^2(x)
x^{-2\sigma-1}dx, ~~~~0<\sigma<1, \eqno(2.24)$$
may then be found by appealing to Parseval's relation for Mellin transforms.

Here, we give a direct derivation of (2.24) from (2.23) by using a Dirac delta function
identity (A.14) of the Appendix.  We have
$$\int_R \left|{{\zeta(s)} \over s}\right|^{2n}dt=\int_R\int_0^\infty \int_0^\infty {{\varphi_n(x_1)} \over x_1^{s+1}}{{\varphi_n(x_1)} \over x_2^{s^*+1}}dx_1dx_2dt$$
$$=\int_R\int_0^\infty \int_0^\infty {{\varphi_n(x_1)\varphi(x_2)} \over{(x_1x_2)^{\sigma+1}}}
\left({x_2 \over x_1}\right)^{it} dx_1dx_2dt$$
$$=\int_R\int_0^\infty \int_0^\infty {{\varphi_n(x_1)\varphi(x_2)} \over{(x_1x_2)^{\sigma+1}}}
\cos\left[t\ln\left({x_2 \over x_1}\right)\right] dx_1dx_2dt$$
$$=2\pi\int_0^\infty \int_0^\infty {{\varphi_n(x_1)\varphi(x_2)} \over {(x_1x_2)^{\sigma+1}}} \delta\left[\ln\left({x_2 \over x_1}\right)\right]dx_1dx_2$$
$$=2\pi\int_0^\infty \int_0^\infty {{\varphi_n(x_1)\varphi(x_2)} \over {(x_1x_2)^{\sigma+1}}} x_1\delta(x_1-x_2)dx_1dx_2$$
$$=2\pi\int_0^\infty {{\varphi^2_n(x)} \over x^{2\sigma+1}}dx.  \eqno(2.25)$$

We may note that Lemma 1 affords the evaluation of all integrals
$$\int_0^\infty {{\{x\}^n} \over x^{2\sigma+1}}dx=\int_0^1 y^n \zeta(2\sigma+1,y)dy,
\eqno(2.26)$$
via $n$ integrations by parts.

{\it \bf Proposition 8}.  Let $_2F_1$ be the Gauss hypergeometric function (e.g., 
\cite{nbs,grad}).  (a) We have
$$\int_0^\infty {{\{x\}} \over x}e^{-x/T}dx=\sum_{\ell=0}^\infty \int_\ell^{\ell+1} {{(x-\ell)} \over x}e^{-x/T}dx=\sum_{\ell=0}^\infty \int_0^1 {y \over {y+\ell}} e^{(-y+\ell)/T}dy$$   
$$=\int_0^1 e^{-y/T} ~_2F_1\left(1,y;y+1;e^{-1/T}\right)dy \eqno(2.27)$$
$$={1 \over {(1-e^{-1/T})}}\int_0^1 e^{-y/T} ~_2F_1\left(1,1;y+1;{1 \over {1-e^{1/T}}}\right)dy.
\eqno(2.28)$$
Relation (2.27) follows immediately from the series definition of $_2F_1$ and the ratio of
Pochhammer symbols $(y)_\ell/(y+1)_\ell=y/(y+\ell)$.  The form (2.28) follows from a 
standard transformation formula for $_2F_1$ (\cite{grad}, p. 1043).  Now either 
(2.27) or (2.28) can be expanded in powers of $1/T$ and integrated termwise.  In particular, 
for (2.27) the key is to use the following expansion valid for $|z-1|<1$ and $|\mbox{arg}
(1-z)|<\pi$:
$$_2F_1(1,y;1+y;z)=y\sum_{k=0}^\infty {{(y)_k} \over {k!}}[\psi(k+1)-\psi(k+y)-\ln(1-z)]
(1-z)^k, \eqno(2.29)$$
where $(y)_0=1$.  This expansion is the $n=0$ case of (9.7.5) in \cite{lebedev}.
We then put $z=e^{-1/T}$, with $\ln(1-e^{-1/T})=-\ln T-1/2T+1/24T^2 +O(1/T^4)$.
This yields part (a), with for instance the leading term in (1.25) given by
$$\int_0^1 [-\gamma+\ln T-\psi(y)]ydy={1 \over 2}[\ln(2\pi)-\gamma+\ln T].  \eqno(2.30)$$

(b) Now we have
$$\int_0^\infty {{\{x\}^2} \over x^2}e^{-x/T}dx=\sum_{\ell=0}^\infty \int_\ell^{\ell+1} {{(x-\ell)^2} \over x^2}e^{-x/T}dx=\sum_{\ell=0}^\infty \int_0^1 {y^2 \over {(y+\ell)^2}} e^{(-y+\ell)/T}dy$$ 
$$=\int_0^1 y^2 \Phi\left(e^{-1/T},2,y\right)e^{-y/T}dy, \eqno(2.31)$$
where $\Phi(1,2,y)=\zeta(2,y)$.  We again expand in powers of $1/T$ and integrate.  
We specialize an expansion of $\Phi(z,m,a)$ for $m \geq 2$ an integer \cite{sri} (p. 123)
to write
$$z^a\Phi(z,2,a)=\zeta(2,a)+\zeta(0,a){{\ln^2 z} \over 2}+\sum_{k=3}^\infty \zeta(2-k,a)
{{\ln^k z} \over {k!}}+\ln z[1-\gamma-\psi(a)-\ln(\ln(1/z))], \eqno(2.32)$$
where $1-\gamma=\psi(2)$, $\zeta(0,a)=1/2-a$, and more generally we recognize
$\zeta(-m,a)=-B_{m+1}(a)/(m+1)$, with $B_j(x)$ the Bernoulli polynomials.  With $z=e^{-1/T}$
we then have
$$e^{-y/T}\Phi(e^{-1/T},2,y)=\zeta(2,y)+{1 \over T}[\gamma-1+\psi(y)-\ln T]+{{\zeta(0,y)} 
\over {2T^2}} + \sum_{k=3}^\infty {{(-1)^k} \over {k!}} {{\zeta(2-k,y)} \over T^k}.  \eqno(2.33)$$
For the leading term in (1.31) we integrate by parts twice to find
$$\int_0^1 y^2 \zeta(2,y)dy=\int_0^1 y^2 \psi'(y)dy=\ln(2\pi)-\gamma.  \eqno(2.34)$$
Here, the relation (2.18) applies, with $\psi'$ the trigamma function.  For the next term
we use the integral
$$\int_0^1 y^2 \psi(y)dy=-2\int_0^1 y \ln \Gamma(y)dy=2\ln A-{1 \over 2}\ln(2\pi).  \eqno(2.35)$$
This evaluation can be based upon the use of the Fourier series for $\ln \Gamma(y)$ (e.g.,
\cite{grad}, p. 940), and keeping in mind the relations
$$-\sum_{n=2}^\infty {{\ln n} \over n^2}=\zeta'(2)=\zeta(2)(\gamma+\ln(2\pi)-12\ln A).  \eqno(2.36)$$
Similarly, the higher order terms may be found.

As a supplement, we have
\newline{\bf Proposition 11}.  As $T \to \infty$ we have (a)
$$\int_0^\infty {{\{x\}} \over x}(\ln x) e^{-x/T}dx={1 \over 4}\ln^2 T+O(\ln T),  \eqno(2.37)$$
and (b) for $k \geq 1$
$$\int_0^\infty {{\{x\}} \over x}(\ln^k x) e^{-x/T}dx={{(-1)^{k+1}} \over {2(k+1)!}}\ln^{k+1} T+O(\ln^k T).  \eqno(2.38)$$  

{\it Proof}.  We just provide a sketch, as the approach of \cite{ivic2} may be used.
The results may be based upon the Mellin transformation and its inverse
$$\int_0^\infty t^{\alpha-1} e^{-t}\ln t ~dt=\Gamma(\alpha)\psi(\alpha), ~~~~\mbox{Re}~
\alpha>0, \eqno(2.39)$$
and
$$e^{-z}\ln z ={1 \over {2\pi i}}\int_{c-i\infty}^{c+i\infty} z^{-s}\Gamma(s)\psi(s)ds, ~~~~
\mbox{Re}~ z>0, ~~c>0.  \eqno(2.40)$$
and logarithmic differentiations with respect to $\alpha$ thereof.  For instance, (2.40) 
leads to
$${1 \over {2\pi i}}\int_{c-i\infty}^{c+i\infty} {{\zeta(s)} \over s} T^s \Gamma(s)\psi(s)
ds=-\int_0^\infty {{\{x\}} \over x}(\ln x)e^{-x/T}dx, ~~~~0<c<1,  \eqno(2.41)$$
where $\psi(s)$ has a simple pole at the origin, $\psi(s)=-1/s-\gamma +\zeta(2) s +O(s^2)$.  The dominant contribution comes from the origin and computing the residue there gives part (a).  
For part (b), the leading contribution comes from the residue at $s=0$ of
${{\zeta(s)} \over s} T^s \Gamma(s)[\psi(s)]^k$ and this is easily seen to be as given.

{\it Remark}.  In relation to part (a), we have
$$\int_0^\infty {{\{x\}} \over x}(\ln x) e^{-x/T}dx=\sum_{\ell=0}^\infty \int_\ell^{\ell+1} {{(x-\ell)} \over x}(\ln x) e^{-x/T}dx=\sum_{\ell=0}^\infty \int_0^1 {y \over {y+\ell}} \ln(y+\ell) e^{(-y+\ell)/T}dy$$
$$=-\int_0^1 ye^{-y/T} \left.{\partial \over {\partial s}}\right|_{s=1} \Phi\left(e^{-1/T},s,
y \right)dy.  \eqno(2.42)$$
For $s \notin N$, it follows that \cite{sri} (p. 123)
$$z^a \partial_s\Phi(z,s,a)=-\psi(1-s)\Gamma(1-s)\left(\ln{1 \over z}\right)^{s-1}+\Gamma(1-s)
\left(\ln{1 \over z}\right)^{s-1}\ln\left[\ln\left({1 \over z}\right)\right]$$
$$+\sum_{k=0}^\infty \zeta'(s-k,a){{\ln^k z} \over {k!}}.  \eqno(2.43)$$
A limit result of this equation may be useful in developing the large-$T$ asymptotic form of
(2.42).

{\it \bf Proposition 9}.  We use the representation (2.23) and apply the inverse Mellin
transformation to write
$$\varphi_n(s)={1 \over {2\pi i}}\int_{\sigma >0} {{\zeta^n(s)} \over {(-s)^n}} x^s ds.
\eqno(2.44)$$
We shift the line of integration slightly to the left of the imaginary axis and evaluate
the residue at the origin.  Within the integrand
$${{\zeta^n(s)} \over {(-s)^n}} x^s = {{\zeta^n(s)} \over {(-s)^n}} \sum_{j=0}^\infty
{s^j \over {j!}} \ln^j x, \eqno(2.45)$$
we recall that $\zeta(0)=-1/2$ and $\zeta'(0)=-(1/2)\ln (2\pi)$, and the result follows.
By further using the multinomial theorem, additional refinements could be included.

{\it \bf Proposition 10}.  We introduce the cosine integral
$$\mbox{Ci}(z) \equiv -\int_z^\infty {{\cos t} \over t}dt.  \eqno(2.46)$$
The function $\varphi_2$ may be decomposed in several ways.  Here it suffices to write
$$\varphi_2(x)=I_1(x)+I_\infty(x)=\left(\int_0^1+\int_1^\infty\right)\{u\}\left\{ {x \over u}
\right\} {{du} \over u}$$
$$=\int_0^1 \left\{ {x \over u}\right\} du+\int_1^\infty \{u\}\left\{ {x \over u} \right\} 
{{du} \over u}. \eqno(2.47)$$

We have
\newline{\bf Lemma 2}.  For $x$ not an integer, we have (a)
$$I_1(x)=P_1(x)+{1 \over 2}+2x \sum_{j=1}^\infty \mbox{Ci}(2\pi j x).  \eqno(2.48)$$
For $x$ an integer, the $P_1$ term in this equation should be omitted.  (b) For $k$ an
integer we have
$$I_1(k)=k\psi(k)-k\ln k+1.  \eqno(2.49)$$

We note that $I_1(1)=1-\gamma$ and as $k \to \infty$, $I_1(k)=1/2-1/12k+1/120k^3+O(1/k^5)$.
Therefore, $\lim_{k \to \infty} I_1(k)=1/2$.

{\it Proof}.  (a) We have the Fourier series (\cite{nbs}, p. 805)
$$P_1(x)+{1 \over 2}=\{x\}={1 \over 2}-\sum_{j=1}^\infty {{\sin(2\pi jx)} \over {\pi j}},
\eqno(2.50)$$
holding for $x$ not an integer.  Then
$$I_1(x)=\int_1^\infty {{\{xy\}} \over y^2}dy={1 \over 2}-{1 \over \pi}\sum_{j=1}^\infty
{1 \over j}\int_1^\infty {{\sin(2\pi j xy)} \over y^2}dy \eqno(2.51)$$
$$={1 \over 2}-{1 \over \pi}\sum_{j=1}^\infty {1 \over j}[\sin(2\pi jx)-2\pi j x \mbox{Ci}
(2\pi jx)],  \eqno(2.52)$$
by integrating by parts.  The integral of (2.51) is absolutely convergent and the 
interchange of summation and integration is justified there.  Applying (2.50) again gives 
the expression (2.48)

For (b) we have
$$\sum_{j=1}^\infty \mbox{Ci}(2\pi j k) = -\sum_{j=1}^\infty \int_{2\pi j k}^\infty
{{\cos t} \over t}dt=-\sum_{j=1}^\infty \int_0^\infty {{\cos(v+2\pi j k)} \over {v+2\pi j k}}
dv$$
$$=-\sum_{j=1}^\infty \int_0^\infty \cos v \int_0^\infty e^{-(v+2\pi j k)x} dxdv$$
$$=-\int_0^\infty \int_0^\infty {{e^{-vx} \cos v} \over {(e^{2\pi kx}-1)}} dv dx$$
$$=-\int_0^\infty {x \over {(e^{2\pi k x}-1)}} {{dx} \over {(1+x^2)}}$$
$$={1 \over 2}\psi(k)-{1 \over 2}\ln k+{1 \over {4k}}.  \eqno(2.53)$$
The last integral has been evaluated by using (\cite{sri}, p. 91).

We have
$$I_\infty(x)=\int_0^1 \left\{{1 \over y}\right\}\{xy\}{{dy} \over y}.  \eqno(2.54)$$
Therefore, we have the well known integral
$$I_\infty(1)=\int_0^1 \left\{{1 \over y}\right\}\{y\}{{dy} \over y}
=\int_0^1 \left\{{1 \over y}\right\}dy=\int_1^\infty {{\{t\}} \over t^2}dt=1-\gamma.  \eqno(2.55)$$
Thus $I_1(1)=I_\infty(1)=1-\gamma$ for part (a) of the Proposition.

For part (b) we have
$$I_\infty(2)=\int_0^1 \left\{{1 \over y}\right\} \{2y\}{{dy} \over y}$$
$$=\left(\int_0^{1/2}+\int_{1/2}^1\right) \left\{{1 \over y}\right\} \{2y\}{{dy} \over y}$$
$$=2\int_2^\infty \{u\} {{du} \over u^2}+\int_1^2 \{u\} (2-u){{du} \over u^2}$$
$$=2\sum_{j=2}^\infty \int_j^{j+1} {{(u-j)} \over u^2}du+\int_1^2 (u-1)(2-u) {{du} \over u^2}$$
$$=2\sum_{j=2}^\infty \left[\ln\left({{j+1} \over j}\right)-{1 \over {j+1}}\right]-2+3\ln 2$$
$$=2\left[{3 \over 2}-\gamma-\ln 2\right]-2+3\ln 2 = -2\gamma+1+\ln 2.  \eqno(2.56)$$
Together with Lemma 2 we have part (b) of the Proposition.

In conjunction with more general considerations of $I_\infty(k)$ with $k$
a positive integer, we have the integral
$$\int_k^\infty \{u\}{{du} \over u^2}=\sum_{j=k}^\infty \int_j^{j+1} (u-j) {{du} \over u^2}$$
$$=\sum_{j=k}^\infty\left[\ln\left({{j+1} \over j}\right)-{1 \over {j+1}}\right]
=H_k-\gamma - \ln k, \eqno(2.57)$$
where $H_k \equiv \sum_{\ell=1}^k 1/\ell$ is the $k$th harmonic number.  With the aid of (2.57), we find $I_\infty(3)=1/2-3\gamma+\ln(27/4)$, and then from Lemma 2 we obtain 
part (c).  Similarly, $I_\infty(4)=-1/3-4\gamma+\ln(256/9)$, 
$I_\infty(5)=-17/12-5\gamma+\ln(5^8/2304)$, $I_\infty(6)=-27/10-6\gamma+\ln(26244/25)$, 
$I_\infty(7)=-83/20-7\gamma+2\ln(7^7/8640)$, $I_\infty(8)=-201/35-8\gamma +\ln(2^{36}/893025)$, $I_\infty(9)=-2089/280-9\gamma-24\ln 2+34\ln 3-2\ln 35$,
and $I_\infty(10)=-2341/253-10\gamma-\ln 2-12 \ln 3+21\ln 5  -2\ln 7$, and the remaining 
parts follow.

\medskip
\centerline{\bf Acknowledgement}
\medskip

Useful discussions with J. Bernard are gratefully acknowledged.  

\pagebreak
\centerline{\bf Appendix:  Evaluations of $\int_R \left|{{\zeta(s)} \over s}\right|^2 dt$}

For $\sigma>1$ we have
$$\int_{-\infty}^\infty \left|{{\zeta(s)} \over s}\right|^2 dt=\int_{-\infty}^\infty
\left|\sum_{n=1}^\infty {1 \over n^{\sigma+it}}\right|^2 {{dt} \over {\sigma^2+t^2}}$$
$$=\sum_{n,m \geq 1}{1 \over {(nm)^\sigma}}\int_{-\infty}^\infty \left({m \over n} \right)^{it} {{dt} \over {\sigma^2+t^2}}$$
$$=\sum_{n,m \geq 1}{1 \over {(nm)^\sigma}}\int_{-\infty}^\infty \cos\left[t\ \ln \left({m \over n}\right)\right] {{dt} \over {\sigma^2+t^2}}$$
$$={\pi \over \sigma}\sum_{n=1}^\infty {1 \over n^{2\sigma}}+\sum_{n \geq 1} \sum_{\stackrel
{m \geq 1}{m \neq n}}{1 \over {(nm)^\sigma}}\int_{-\infty}^\infty \cos\left[t\ \ln \left({m \over n} \right)\right] {{dt} \over {\sigma^2+t^2}}$$
$$={\pi \over \sigma}\left[\zeta(2\sigma)+2 \sum_{m=1}^\infty \sum_{n=1}^{m-1} {1 \over m^{2\sigma}}\right]={\pi \over \sigma}\left[\zeta(2\sigma)+2 \sum_{m=1}^\infty {{m-1} \over m^{2\sigma}}\right]$$
$$={\pi \over \sigma}[\zeta(2\sigma)+2\zeta(2\sigma-1)-2\zeta(2\sigma)]
={\pi \over \sigma}[2\zeta(2\sigma-1)-\zeta(2\sigma)].   \eqno(A.1)$$

We now provide an alternative evaluation, through the $a=1$ special case of the representation
$$\zeta(s,a)={a^{-s} \over 2}+{a^{1-s} \over {s-1}}-s\int_0^\infty {{P_1(x)} \over
{(x+a)^{s+1}}} dx, ~~~~\mbox{Re} ~s >-1, \eqno(A.2)$$
where $P_1(x)=x-[x]-1/2=\{x\}-1/2$ is the first periodic Bernoulli polynomial, and
$\{x\}$ denotes the fractional part of $x$.  I.e., we form $|\zeta(s)|^2=\zeta(s)\zeta^*(s)
=\zeta(s)\zeta(s^*)$ with
$$\zeta(s^*)={1 \over 2}+{1 \over {s^*-1}}-s^*\int_1^\infty {{P_1(x)} \over
{x^{s^*+1}}} dx, ~~~~\sigma >-1. \eqno(A.3)$$ 
We first describe the contributions
$${1 \over 2}\int_R \left({1 \over {s-1}}+{1 \over {s^*-1}}\right){{dt} \over {|s|^2}}
={{\pi(\sigma-1)} \over {(2\sigma-1)}}\left({1 \over {\sigma-1}}-{1\over \sigma}\right),
\eqno(A.4)$$
$${1 \over 4}\int_R {{dt} \over {|s|^2}}-{1 \over 2}\int_R \int_1^\infty P_1(x)\left({s
\over x^{s+1}}+{s^* \over x^{s^*+1}}\right)dx {{dt} \over {|s|^2}}={\pi \over {2\sigma}}
\left[\zeta(2\sigma)-{1 \over {(2\sigma-1)}}\right],  \eqno(A.5)$$
and
$$\int_R {{dt} \over {|s|^2|s-1|^2}}=\int_{-\infty}^\infty {{dt} \over {(\sigma^2+t^2)
[(\sigma-1)^2+t^2]}}={\pi \over {2\sigma-1}}\left(-{1 \over \sigma}+{1 \over {\sigma-1}}
\right).  \eqno(A.6)$$

For the integrand of (A.4) we use 2Re$\left({1 \over {s-1}}\right)$ and partial fractions so that
$${1 \over 2}\int_R \left({1 \over {s-1}}+{1 \over {s^*-1}}\right){{dt} \over {|s|^2}}
=(\sigma-1)\int_R {1 \over {(\sigma^2+t^2)}}{{dt} \over {[(\sigma-1)^2+t^2]}}$$
$$={{(\sigma-1)} \over {(2\sigma-1)}} \int_R \left[-{1 \over {\sigma^2+t^2}}+{1 \over 
{(\sigma-1)^2+t^2}}\right]dt={{(\sigma-1)} \over {(2\sigma-1)}}\pi \left(-{1\over \sigma} +{1\over {\sigma-1}}\right).  \eqno(A.7)$$
For (A.5) we have 2Re$\left({s \over x^{s+1}}\right)={2 \over x^{\sigma+1}}[\sigma \cos(t
\ln x)+t \sin(t\ln x)]$ and have from (A.2) that
$$-\int_1^\infty {{P_1(x)} \over x^{2\sigma+1}}dx={1 \over {2\sigma}}\left[\zeta(2\sigma)-{1 
\over 2}-{1 \over {(2\sigma-1)}}\right].  \eqno(A.8)$$

We next have the terms
$$-\int_R \int_1^\infty P_1(x)\left[{s \over {(s^*-1)x^{s+1}}}+{s^* \over {(s-1)x^{s^*+1}}}
\right]dx {{dt} \over {|s|^2}}$$
$$=4\pi\left\{{1 \over{(2\sigma-1)}}\left[\zeta(2\sigma-1)-{1 \over 2}-{1 \over {2(\sigma-1)}}
\right]-{1 \over {2\sigma}}\left[\zeta(2\sigma)-{1 \over 2}-{1 \over {(2\sigma-1)}}\right]
\right\}.  \eqno(A.9)$$
Here,
$$-\int_R \int_1^\infty P_1(x)\left[{s \over {(s^*-1)x^{s+1}}}+{s^* \over {(s-1)x^{s^*+1}}}
\right]dx {{dt} \over {|s|^2}}=-2\int_R \int_1^\infty P_1(x)\mbox{Re}\left[{s \over {(s^*-1)x^{s+1}}}\right]dx{{dt} \over {|s|^2}}$$
$$=-2\int_1^\infty {{P_1(x)} \over x^{\sigma+1}}{{\left\{[\sigma(\sigma-1)-t^2]\cos(t\ln x)+(2\sigma-1)t\sin(t\ln x)\right\}} \over {[(\sigma-1)^2 +t^2](\sigma^2+t^2)}} dt$$
$$=-4\pi \int_1^\infty {{P_1(x)} \over x^{2\sigma}}
\left[1-{1 \over x}\right] dx.  \eqno(A.10)$$
Using (A.8) then gives (A.9).  In obtaining (A.10) we have used the integrals via partial
fractions
$$-\int_{-\infty}^\infty {{t^2 \cos(t\ln x)} \over {(\sigma^2+t^2)[(\sigma-1)^2+t^2]}}dt
=-{\pi \over {(2\sigma-1)}}{1 \over x^{\sigma-1}}\left[{\sigma \over x}-(\sigma-1)\right],
\eqno(A.11a)$$
and
$$(2\sigma-1)\int_{-\infty}^\infty {{t \sin(t\ln x)} \over {(\sigma^2+t^2)[(\sigma-1)^2+t^2]}} dt ={\pi \over x^{\sigma-1}}\left(-{1 \over x}+1\right).  \eqno(A.11b)$$

Finally, we have a term
$$\int_R \int_1^\infty \int_1^\infty {{P_1(x_1)P_1(x_2)} \over {(x_1x_2)^{\sigma+1}}}
\left({x_2 \over x_1}\right)^{it}dx_1dx_2dt=2\pi \int_1^\infty {{P_1^2(x)} \over 
x^{2\sigma+1}}dx.  \eqno(A.12)$$
Here we have used 
$$\int_R \cos[t\ln(x_2/x_1)]dt=2\pi\delta[\ln(x_2/x_1)]=2\pi x_1\delta(x_1-x_2), \eqno(A.13)$$ with $\delta$ the Dirac delta function.  The second equality in (A.13) follows from the
identity
$$\delta[f(x)]=\sum_i {{\delta(x-x_i)} \over {|df/dx|_{x=x_i}}}, \eqno(A.14)$$
where the $x_i$ are the roots of the function $f$ in the interval of integration.

The evaluation of (A.12) proceeds by using integration by parts:
$$\int_1^\infty {{P_1^2(x)} \over x^{2\sigma+1}}dx=-{1 \over {2\sigma}}\int_1^\infty
P_1^2(x) \left({d \over {dx}}{1 \over x^{2\sigma}}\right)dx$$
$$={1 \over \sigma}\int_1^\infty {{P_1(x)} \over x^{2\sigma}}\left[1-\sum_j \delta(x-j)\right]dx +{1 \over {8\sigma}}$$
$$={1 \over {\sigma(2\sigma-1)}}\left[-\zeta(2\sigma-1)+{1 \over 2}+{1 \over {2(\sigma-1)}}\right]+{1 \over {8\sigma}}+{1 \over {2\sigma}}\sum_{j=2}^ \infty {1 \over j^{2\sigma}}$$
$$={1 \over {\sigma(2\sigma-1)}}\left[-\zeta(2\sigma-1)+{1 \over 2}+{1 \over {2(\sigma-1)}}\right]+{1 \over {8\sigma}}+{1 \over {2\sigma}}[\zeta(2\sigma)-1].  \eqno(A.15)$$

A second evaluation of (A.12) uses Lemma 1.  We have, also using (A.8),
$$\int_1^\infty {{P_1^2(x)} \over x^{2\sigma+1}}dx=\int_1^\infty {{(\{x\}^2-\{x\}+1/4)} \over
x^{2\sigma+1}}dx$$
$$=\int_0^1 y^2 \zeta(2\sigma+1,y+1)dy-\int_1^\infty {{\{x\}} \over x^{2\sigma+1}}dx+{1 \over
{8\sigma}}$$
$$={1 \over {2(\sigma-1)}}-{{\zeta(2\sigma)} \over {2\sigma}}-{{\zeta(2\sigma-1)} \over
{\sigma(2\sigma-1)}}+{1 \over {2\sigma}}\left[\zeta(2\sigma)-1-{1 \over {2\sigma-1}}\right]
+{1 \over {8\sigma}}$$
$$=-{1 \over {\sigma(2\sigma-1)}}\left[\zeta(2\sigma-1)+{1 \over 2}\right]+{1 \over {2(\sigma-1)}}-{3 \over {8\sigma}}$$
$$={1 \over {2(\sigma-1)}}-{{\zeta(2\sigma-1)} \over {\sigma(2\sigma-1)}}-{1 \over {(2\sigma-1)}}+{1 \over {8\sigma}}.  \eqno(A.16)$$
The integral of (A.12) itself converges for $\sigma>0$.

For a third evaluation of (A.12) we first record without proof the following elementary
sums.
\newline{\bf Lemma 3}.  We have
$$\sum_{j=1}^\infty \left[{a \over {(j+1)^{2\sigma}}}-{b \over j^{2\sigma}}\right] =(a-b)\zeta(2 \sigma)-a, \eqno(A.17a)$$
$$\sum_{j=1}^\infty j\left [{a \over {(j+1)^{2\sigma}}}-{b \over j^{2\sigma}}\right]
=-a\zeta(2 \sigma)+(a-b)\zeta(2\sigma-1), \eqno(A.17b)$$
and
$$\sum_{j=1}^\infty j^2 \left[{a \over {(j+1)^{2\sigma}}}-{b \over j^{2\sigma}}\right] =a\zeta(2 \sigma)-2a\zeta(2\sigma-1)+(a-b)\zeta(2\sigma-2). \eqno(A.17c)$$

Then
$$\int_1^\infty {{P_1^2(x)} \over x^{2\sigma+1}}dx=\int_1^\infty {{(x-[x]+1/2)^2} \over
x^{2\sigma+1}}dx$$
$$=\int_1^\infty {{(x^2+[x]^2+1/4-2[x]x-x+[x])} \over x^{2\sigma+1}}dx$$
$$=\sum_{j=1}^\infty \int_j^{j+1} {{(x^2+j^2+1/4-2jx-x+j)} \over x^{2\sigma+1}}dx$$
$$={1 \over {8\sigma(\sigma-1)(2\sigma-1)}}\sum_{j=1}^\infty \left \{[1-3\sigma+2\sigma^2
+4(1-\sigma)j+4j^2]{1 \over j^{2\sigma}} \right.$$
$$\left. -[1+\sigma+2\sigma^2+4(1+\sigma)j+4j^2]{1 \over {(j+1)^{2\sigma}}} \right\}.  \eqno(A.18)$$
Then the application of Lemma 3 yields (A.16).

\pagebreak

\end{document}